\documentclass[aps,prl,amsmath,reprint,superscriptaddress]{revtex4-2}
\usepackage{bm}
\usepackage{graphicx}
\usepackage{float}
\usepackage{color,soul}
\usepackage[dvipsnames,table]{xcolor}
\usepackage[colorlinks=true, linkcolor=blue, citecolor=blue, urlcolor=blue,breaklinks=true]{hyperref}
\newcommand{\ket}[1]{|#1\rangle}             
\newcommand{\bra}[1]{\langle#1|}             

\colorlet{goodcolor}{green!70!white!30!}
\colorlet{badcolor}{red!70!white!30!}
\makeatletter
\renewcommand{\@caption@fignum@sep}{\ ${\bm |}$\ }%
\makeatother

\newcommand{\PreserveBackslash}[1]{\let\temp=\\#1\let\\=\temp}
\newcolumntype{C}[1]{>{\PreserveBackslash\centering}p{#1}}
\newcolumntype{R}[1]{>{\PreserveBackslash\raggedleft}p{#1}}
\newcolumntype{L}[1]{>{\PreserveBackslash\raggedright}p{#1}}

\renewcommand{\figurename}{\textbf{Fig.}}
\renewcommand{\thefigure}{{\bf \arabic{figure}}}

\begin{document}
	
	\title{Quantum channel correction outperforming direct transmission}

	\author{Sergei Slussarenko}
	\email{s.slussarenko@griffith.edu.au}
	\author{Morgan M. Weston}
	\affiliation{Centre for Quantum Dynamics, Griffith University, Brisbane, Queensland 4111, Australia}
	\affiliation{Centre for Quantum Computation and Communication Technology (Australian Research Council), Australia}
	\author{Lynden K. Shalm}
	\author{Varun B. Verma}
	\author{Sae-Woo Nam}
	\affiliation{National Institute of Standards and Technology, 325 Broadway, Boulder, Colorado 80305, USA.}
	\author{Sacha Kocsis}
	\affiliation{Centre for Quantum Dynamics, Griffith University, Brisbane, Queensland 4111, Australia}
	\affiliation{Current address: Centre  for  Quantum  Computation  and  Communication  Technology,  The University  of  New  South  Wales,  Sydney 2052,  Australia}
	\author{Timothy C. Ralph}
	\email{ralph@physics.uq.edu.au}
	\affiliation{School of Mathematics and Physics, University of Queensland, Brisbane 4072, Australia}
	\affiliation{Centre for Quantum Computation and Communication Technology (Australian Research Council), Australia}
	\author{Geoff J. Pryde}
	\email{g.pryde@griffith.edu.au}
	\affiliation{Centre for Quantum Dynamics, Griffith University, Brisbane, Queensland 4111, Australia}
	\affiliation{Centre for Quantum Computation and Communication Technology (Australian Research Council), Australia}
	\begin{abstract}
	Long-distance optical quantum channels are necessarily lossy, leading to errors in transmitted quantum information, entanglement degradation and, ultimately, poor protocol performance. Quantum states carrying information in the channel can be probabilistically amplified to compensate for loss, but are destroyed when amplification fails. Quantum correction of the channel itself is therefore required, but break-even performance---where arbitrary states can be better transmitted through a corrected channel than an uncorrected one---has so far remained out of reach. Here we perform distillation by heralded amplification to improve a noisy entanglement channel. We subsequently employ entanglement swapping to demonstrate that arbitrary quantum information transmission is unconditionally improved---i.e. without relying on postselection or post-processing of data---compared to the uncorrected channel. In this way, it represents realisation of a genuine quantum relay. Our channel correction for single-mode quantum states will find use in quantum repeater, communication and metrology applications.    
	\end{abstract}
	
	\maketitle
	\setlength{\parskip}{0cm plus1mm minus1mm}
	\raggedbottom
\section{Introduction}
	Loss-induced noise, e.g.\ from scattering and diffraction, is inevitable in long-distance information transfer. Although some communication schemes permit postprocessing of measurement results to discard vacuum or noise, high-demand applications such as device-independent protocols~\cite{acin07}, distributed quantum computing~\cite{rev_flamini18,rev_slussarenko19}, or quantum metrology schemes~\cite{gottesman12} cannot rely on postselection. Such applications require genuine entanglement distillation schemes to overcome the effects of loss. Therefore, fully heralded protocols that can demonstrably show an advantage in transmitting states over a noise-reduced entanglement channel are critically important. This is the quantum communication equivalent of surpassing the break-even point for quantum error correction~\cite{ofek16}.
	
	Photon linear loss reduces the amplitude of the creation operator associated with a particular mode (channel) according to a linear transformation, and adds noise by mixing in one or more other optical modes in the process. This process is gaussian, mapping one gaussian distribution onto another. Loss-induced noise affects any type of optical encoding. It provides a particular challenge for coherent and squeezed states, as it has been proven that it is impossible to correct gaussian noise on gaussian states using only gaussian operations~\cite{niset09}. 
	
	Techniques for addressing loss-induced noise broadly include entanglement-swapping-based quantum nondemolition style measurements~\cite{pan98,kaltenbaek09,weston18,tsujimoto20} and their extensions~\cite{abdelkhalek16,chen17n,xu17,li19}, distillation techniques based on noiseless linear amplification~\cite{xiang10,ferreyrol10,micuda12,ulanov15}, and photonic qubit precertification~\cite{meyerscott16}. Among these, schemes based on noiseless linear amplification (or heralded amplification, HA) have shown great potential in amplifying single-mode~\cite{xiang10} or two-mode~\cite{kocsis13,bruno16} qubits, correcting for the effects of loss on entangled states~\cite{xiang10,ulanov15,monteiro17}, and in conditionally exceeding quantum cloning bounds~\cite{haw16}.
 
\begin{figure*}[t]
		\includegraphics{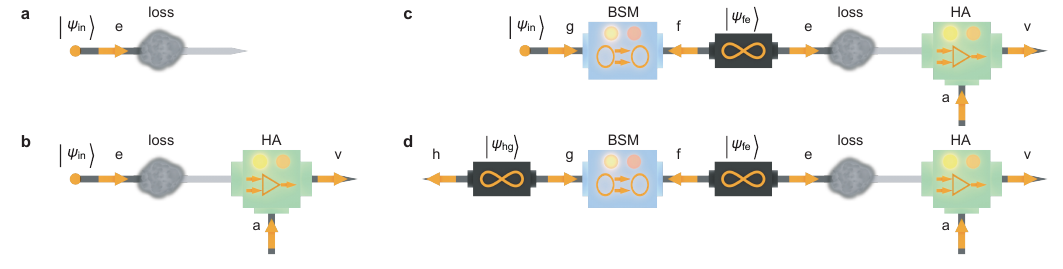}
		\caption{ \label{fig1} \textbf{Conceptual representations of quantum state transmission through a lossy channel, with and without correction.}  A quantum state, here a qubit encoded in a single mode ``e'', is transmitted through a lossy channel, which degrades the state quality. \textbf{b} After the loss, the noisy state can be corrected using a heralded amplifier (HA). Mode ``a'' carries the ancilla photon that powers the HA. The operation  of HA has an independent success signal, so postselection is not required. However, HA failure destroys the state. \textbf{c} By adding a mode-entangled state $\ket{\psi_\textrm{fe}}$, success of the HA heralds a noise-corrected quantum channel. This can be used upon success by teleporting a qubit in mode ``g'' onto mode ``v'' via a Bell state measurement (BSM) between ``g'' and ``f''. {\bf d} Instead of transmitting a qubit $\ket{\psi_\textrm{in}}$ through the lossy or corrected channel, it is possible to transmit half of an entangled state, leading to distributed entanglement through a heralded corrected channel in the last case.}
	\end{figure*}
 
	In the absence of error correcting codes, noise reduction in quantum mechanics is necessarily probabilistic and one needs to consider how this affects the noise mitigation process. Broadly, noise reduction or correction can be grouped into three categories: postselection, where final detection of the (probabilistically) corrected state identifies whether correction was successful; heralded state correction, where an independent signal flags (or heralds) noise reduction on the state without need for postselection but where the state is destroyed upon failure; and heralded channel correction, where successful heralding prepares a channel that can be used to transmit a state, and one can refrain from sending the state if channel noise reduction fails. Each of these tasks is more challenging than the previous. 
	
	Here, we experimentally implement heralded channel correction using HA~\cite{xiang10}, use it to distribute entanglement via entanglement swapping~\cite{pan98} and demonstrate reduced loss-induced errors in a quantum channel~\cite{ralph11}. Our scheme thus demonstrates the core task of a quantum relay in distributing quantum information over a channel affected by loss.
\section{Results}
\subsection{Theory}
	From an information perspective, the channel is represented by an entangled subsystem of quantum optical modes. Such entanglement, once corrected for loss, can be used to implement a quantum teleportation scheme to transfer quantum information across the channel. The specific type of mode entanglement we generate for this channel is obtained by splitting a single photon between two modes. This single-rail encoding~\cite{lund02} is of current interest~\cite{loredo19,caspar20} due to its applications in twin-field quantum key distribution protocols~\cite{minder19} and recently demonstrated techniques for converting between single rail and dual rail encoding~\cite{drahi21}.
 
\begin{figure*}
		\includegraphics{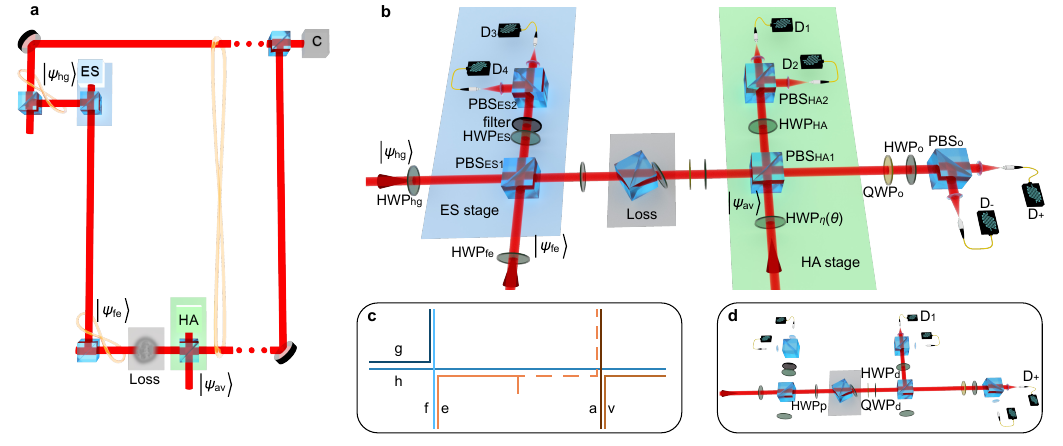}
		\caption{ \label{fig2} {\bf Concept and layout of the experimental apparatus.} \textbf{a} Concept of the experiment, which implements the scheme of Fig.~\ref{fig1}\textbf{d}. Half of the entangled state $\ket{\psi_\textrm{hg}}$ is teleported through a heralded corrected channel. This comprises an entangled source, loss on one mode (the distribution mode), and an HA. This leads to entanglement between modes ``h'' and ``v'' that, ideally, is larger than the entanglement between ``f'' and ``e'' after ``e'' goes through the lossy channel. Modes ``h'' and ``v'' are then brought together for a joint measurement to determine the concurrence $C$. {\bf b} Experimental setup for the error-corrected quantum channel. This diagram has a one-to-one mode mapping to part \textbf{a}, as shown in subfigure \textbf{c}, but some of the modes are now spatially overlaid and distinguished by polarization. Blue and green backgrounds highlight ES and HA stages, respectively. Grey background highlights polarization dependent loss.	Optical axes of ${\rm HWP}_{\rm fe}$ and ${\rm HWP}_{\rm hg}$, set at $\pi/8$ with respect to the horizontal axis, prepared $\ket{\psi_{\rm fe}}$ and $\ket{\psi_{\rm hg}}$ states, respectively. Optical modes input to PBS$_{\rm HA2}$ and PBS$_{\rm ES2}$ were mixed by ${\rm HWP}_{\rm HA}$ and ${\rm HWP}_{\rm ES}$, set at $\pi/8$. ${\rm HWP}_{\rm \eta}(\theta)$, with optic axis orientated at angle $\theta$, initialized the HA resource state $\ket{\psi_{\rm av}}$, with $\eta=\sin{(2\theta)}$. HA success was heralded by a single detection event in either superconducting nanowire single photon detector (SNSPD) $D_1$ or $D_2$ and ES success was heralded by a single detection event in either $D_3$ or $D_4$. Detectors $D_+$ and $D_-$, together with QWP$_{\rm o}$, HWP$_{\rm o}$ and PBS$_{\rm o}$ were used to perform polarization state tomography on the modes ``h'' and ``v''. {\bf c} Mode propagation inside the setup in case of the error-corrected channel. Mode ``h'' propagates through the setup unaffected by loss or other optical components. {\bf d} Same setup but with single photon input, used to test direct transmission through loss to compare the final concurrences. HWP$_{\rm p}$ was set to $\pi/8$ in order to prepare $\ket{\psi_{\rm fe}}$. QWP$_{\rm d}$, HWP$_{\rm d}$ and PBS$_{\rm HA1}$, together with detectors $D_+$ and $D_1$ were used to perform polarization tomography of the state after the loss. Note that the ultrahigh-heralding-efficiency photon sources required for this protocol are not shown in the figure---see Methods for details.}
	\end{figure*}
  
	The theoretical concept is shown in Fig.~\ref{fig1} and explained using the language of single-mode qubits, i.e.\ superpositions of the vacuum and single-photon state in a single optical mode. First consider Fig.~\ref{fig1}\textbf{a}, \textbf{b}, \textbf{c}. The aim is to transmit such a qubit $\ket{\psi_{\rm in}}=\alpha\ket{0_{\rm e}}+\beta\ket{1_{\rm e}}$, initially encoded in mode ``e'', as per Fig.~\ref{fig1}\textbf{a}. Transmission of ``e'' through a lossy channel degrades it to
	\begin{equation}
	\rho_{\rm out}= |\beta|^2 (1- T) |0\rangle\langle 0| + (\alpha |0 \rangle +\sqrt{T} \beta |1 \rangle)(\alpha^* \langle 0| + \sqrt{T}\beta^*\langle 1|),
		\label{Eq1}
	\end{equation} 
	where the transmission $T$ can also be written as $1 - L$ with $L$ being the loss. It has been shown that postselection-based filtering can reduce the noise on states of this general kind~\cite{chrzanowski14}, but this is useful for a limited range of applications.  Alternatively, a heralded amplifier (Fig.~\ref{fig1}{\bf b}), applied after the loss, changes the degraded state according to 
	\begin{equation}
	\rho_{\rm HA} \rightarrow(1-P_{\rm amp})\ket{0}\bra{0}\otimes\Pi_{\rm namp}+P_{\rm amp}\rho_{\rm amp}\otimes\Pi_{\rm amp},
	\label{Eq2}
	\end{equation} 
	where $P_{\rm amp}$ is the probability of success of the amplification and $\Pi_{\rm amp}$ ($\Pi_{\rm namp}$) is the projector onto the subspace of where the heralding signal was (not) received. The amplified state $\rho_{\rm amp}$ is given by
	\begin{eqnarray}
	&&\rho_{\rm amp} = \nonumber \\ 
	&&{{|\beta|^2 (1- T) |0\rangle\langle 0| + (\alpha |0 \rangle + g \sqrt{T} \beta |1 \rangle)(\alpha^* \langle 0| + g \sqrt{T} \beta^*\langle 1|)}\over{1+T |\beta|^2(g^2-1)}},\nonumber \\
		\label{Eq3}
	\end{eqnarray} 
	 where $g\geq1$ is the (amplitude) amplification gain. If $g=1$, the HA acts as a single-mode teleportation stage without amplification.
	
	Although the heralded amplifier can be used to mitigate the effect of loss on the quantum state in the sense that the average fidelity of Eq.~\ref{Eq3} with the initial state exceeds that of Eq.~\ref{Eq1}, the fidelity increase is strictly limited (see Supplementary Section 2). In addition, by itself it is not sufficient to correct for errors in a quantum channel, as information will still be lost in cases when amplification did not succeed. An alternative approach is to use a two-mode entangled state 
	\begin{equation}
	\ket{\psi_{\rm fe}}=\frac{\ket{1_{\rm f}0_{\rm e}}+\ket{0_{\rm f}1_{\rm e}})}{\sqrt{2}},
	\end{equation} 
	as in Fig.~\ref{fig1}\textbf{c}. Such a state represents an entanglement-based channel as it can act as a resource for single-mode teleportation.
	Once the HA has successfully corrected the mode ``e'' from the effects of loss, the other mode ``f'' is used to implement a quantum teleportation scheme to teleport the desired quantum information $\ket{\psi_\textrm{in}}$, encoded in a qubit in mode ``g'', to the output mode ``v'' of the HA. Effectively we have a deterministic, error corrected channel. In the limit of high HA gain and using biased entanglement we can in principle approach an error free channel for any finite level of loss \cite{ralph09,micuda12,monteiro17} (see Supplementary Section 2).

\subsection{Channel correction protocol}	
	An alternative but equally interesting scenario---shown in Fig.~\ref{fig1}\textbf{d},
	---is to consider the transmission of half of a two-mode entangled state (rather than $\ket{\psi_\textrm{in}}$) through the channel. This is the situation we implement in this work. The outcome of such a scheme is an entangled state distributed between modes ``h'' and ``v'', that can be used for further information tasks, in principle. Characterizing the success of the channel correction consists of comparing the amount of entanglement between modes ``h'' and ``v'' in the case of Fig.~\ref{fig1}\textbf{d} to the amount of entanglement in case of direct transmission of a mode of an entangled qubit through loss. We note that if the channel is capable of teleporting entanglement, it can also teleport arbitrary quantum states and, hence entanglement swapping tests the quality with which all possible input states of the relevant Hilbert space can be teleported through the channel.
	
	The single-mode heralded amplification (HA) and entanglement swapping (ES) stages can be implemented with the generalized quantum scissors~\cite{pegg98,xiang10}. In this approach, the input state is interfered non-classically on a beam splitter with one of the modes (mode ``a'' in this example) of a resource state $\ket{\psi_{\rm av}}=\sqrt{\eta}\ket{1_{\rm a} 0_{\rm v}}+\sqrt{1-\eta}\ket{0_{\rm a} 1_{\rm v}}$. The success of the operation is conditioned on getting one and only one detection event in one of the two auxiliary detectors. The difference between entanglement swapping and HA operations is the setting $\eta=1/2$ or $\eta<1/2$, respectively, with the nominal intensity gain given by $g^2=(1-\eta)/\eta$ in an ideal case of perfect interference, unit delivery efficiency of $\ket{\psi_{\rm av}}$ and absence of noise. 
	
	For a small state which exists in the $\{0,1\}$ photon number subspace, a one-to-one mapping exists between pairs of entangled orthogonal spatial modes and a photon polarization qubit:
	\begin{equation}
	\ket{\psi_{\rm in}}=\alpha\ket{1_{\rm H}0_{\rm V}}+\beta\ket{0_{\rm H}1_{\rm V}} = \alpha\ket{{\rm H}}+\beta\ket{{\rm V}},
	\end{equation}
	where H and V represent horizontal and vertical polarizations, respectively.
	For such encoding, mode manipulation is implemented via polarizing beam splitters (PBS) and half- and quarter-wave plates (HWP and QWP), instead of beam splitters.
	
	\subsection{Experiment}
	Our experimental realization of the error-correction scheme of Fig.~\ref{fig1}\textbf{d} is shown in Fig.~\ref{fig2}. We use two photon-pair sources, based on group velocity matched spontaneous parametric downconversion (GVM SPDC), to generate the photons required for the scheme (see Methods for details). One photon-pair source was used to prepare the resource state $\ket{\psi_{\rm fe}}=(\ket{{\rm H}_{\rm f}}+\ket{{\rm V}_{\rm e}})/\sqrt{2}$ and the mode-entangled state for entanglement swapping  $\ket{\psi_{\rm hg}}=(\ket{{\rm H}_{\rm h}}+\ket{{\rm V}_{\rm g}})/\sqrt{2}$. Another source was used to prepare a heralded single photon state $\ket{\psi_{\rm av}}$ as the resource that powers the HA.
	
	The task of demonstrating improved entanglement when transmitting over a heralded corrected channel has stringent technological requirements. High-heralding-efficiency, low noise photon sources, high quality quantum interference, and high-efficiency detectors are necessary in order to complete a 4-photon protocol in such a way that demonstrates improvement over a simple transmission of a single photon, even taking into account the loss. We describe the substance of these requirements, and our means for satisfying them---high-efficiency GVM SPDC sources that directly produce spectrally pure photon pairs, together with high-efficiency superconducting nanowire photon detectors---in the Methods.

		\begin{figure}
		\includegraphics{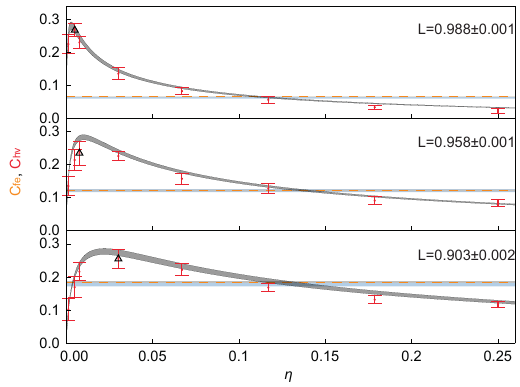}
		\caption{ \label{fig3} {\bf Concurrence measurements of entanglement distributed over different types of channels for three different values of added loss $L$ on mode ``e''.} Orange lines and blue shaded areas correspond to the experimentally measured and theoretically predicted concurrence $C_{\rm fe}$ of $\ket{\psi_{\rm fe}}$ state distributed directly through loss. Red dots and gray shaded areas correspond to the experimentally measured and theoretically predicted concurrence $C_{\rm hv}$ of entanglement distributed via the error-corrected quantum channel Fig.~\ref{fig1}{\bf d}, with the same amount of loss and the same input state $\ket{\psi_{\rm fe}}$. Black triangles highlight data points with maximum observed increase in concurrence. The upper and lower bounds for the theoretical predictions correspond to the lowest and highest observed transmissions for each mode inside the experimental setup, respectively. Error bars and shaded areas on orange lines correspond to the experimentally observed statistical uncertainty of $\pm1$ standard deviation.}
	\end{figure}

	In the $\{\ket{0},\ket{1}\}$ photon number subspace	the density matrix of a two-mode entangled state can be represented as  $\rho=p_{00}\ket{00}\bra{00}+p_{01}\ket{01}\bra{01}+p_{10}\ket{10}\bra{10}+d\ket{01}\bra{10}+d^*\ket{10}\bra{10}+p_{11}\ket{11}\bra{11}$. We assume zero coherence between vacuum, single-photon, and two-photon components of the state. We use concurrence~\cite{chou05} to characterize the entanglement it carries, which in this case is given by 
	\begin{equation}\label{eq:c}
	C=2{\rm Max}[|d|-\sqrt{\rho_{00}\rho_{11}},0].
	\end{equation}
	Its explicit dependence on the vacuum $\rho_{00}$ and high-order $\rho_{11}$ contributions highlights the detrimental effect of loss and higher-order photon emission noise on our scheme. We measure these quantities and $|d|$ by directly measuring the probabilities of detecting vacuum, one- and two-photon component, together with performing a quantum state tomography on output modes of the transmission.
	
	To demonstrate the advantage of our channel error-correction scheme, we first measure the amount of entanglement carried by the state $\ket{\psi_{\rm fe}}$ transmitted through the polarization-dependent loss (as in Fig.~\ref{fig2}{\bf d}). We then compared the result to the amount of entanglement of the same entangled state distributed via the channel error-correction scheme, by measuring the concurrence between modes ``h'' and ``v'', heralded by the corresponding click pattern of the detectors in the HA and ES stages (as in Fig.~\ref{fig1}{\bf d} and Fig.~\ref{fig2}{\bf b}).
	
\begin{figure}
		\includegraphics{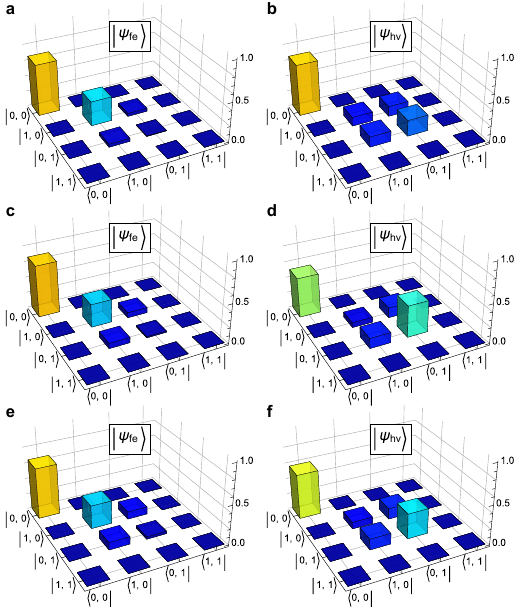}
		\caption{ \label{fig4} {\bf Absolute values of the density matrix elements for quantum states distributed via different types of channels.} {\bf a} Maximally-mode-entangled state sent through an uncorrected quantum channel with $L=0.988\pm0.001$ added loss. {\bf b} Same input state, but distributed through the error-corrected quantum channel with the nominal gain setting from Fig.~\ref{fig3}, at which the observed concurrence was the highest. {\bf c} - {\bf d} and  {\bf e} - {\bf f} -- same as  {\bf a} - {\bf b}, but with $L=0.958\pm0.001$ and $L=0.903\pm0.002$, respectively.}
	\end{figure}

	Our experimental results, together with theoretical predictions, are shown in Fig.~\ref{fig3} for three different values of added loss on mode ``e''.   Our experimental data matches theoretical predictions, derived from a quantum circuit model that includes measured efficiencies and higher order emission from the sources but with no free fit parameters. We attribute the slight reduction of the experimentally observed concurrence to the non-perfect HOM interference at the HA stage, which was not taken into account in the theoretical model. It is worth noting that maximum observed concurrence in the state transmitted through the error-corrected channel does not correspond to the gain setting that recovers the state to $(\ket{\rm H}+\ket{\rm V})/\sqrt{2}$, but to a nominal gain setting at which mode ``e'' is over-amplified, as shown in Fig.~\ref{fig4}.
	
	For the comparison in Fig.~\ref{fig3} between the two different channels to be fair, it is important to take into account the channels' operation rates. Due to the probabilistic nature of SPDC, the probability of generating four photons for the error corrected channel is significantly lower than the probability of generating one photon pair for the direct transmission. This leads to lower raw coincidence counts per second in the former case. For a fair comparison we define the operation rate as the rate of input states prepared and sent through the channel, given successful preparation of the channel. As described in Supplementary Section 3, we find that the operation rate of the error corrected channel is higher or equal to the rate of direct transmission, thus providing us a level playing field for the direct comparison of the concurrences for the two scenarios and showcasing the power of our error corrected channel scheme.
	
	\section{Discussion}
	The challenge of overcoming loss in a quantum channel is central for realizing many quantum technologies, including quantum computation and communication. We have demonstrated error-correction of a quantum channel, in a heralded way. Comparing to a much simpler, direct transmission of a single optical mode through loss, our scheme provides unconditional advantage, despite being experimentally more complex. With our best result we have observed up to a factor of $F=4.1\pm0.3$ increase in concurrence over direct transmission, equivalent to an effective improvement in the channel transmission of $\approx 12.9~{\rm dB}$. In principle, it is possible to implement the entanglement swapping operation only after the HA success signal certifies that the channel is ready for transmission---we do not implement that strict time ordering in this experiment. In our experiment verification of the entanglement utilizes lossless transmission of the ``h'' reference mode from sender to receiver. However, this is done for simplicity and is not fundamental. In the field such characterization could be achieved by interfering the outputs locally with classical optical phase references (known technically as `local oscillators'), followed by appropriate (homodyne) detection. Producing remote, synchronised phase references has been demonstrated in continuous variable and twin-field applications.
	
	\section*{Methods}
	\subsection*{Requirements for experimental advantage}
	
	It is challenging to meet the performance requirements of our scheme in the presence of several inevitable experimental imperfections. 	First, the HA  efficiency is limited by the efficiency of $\ket{\psi_{\rm av}}$ delivery and detection~\cite{berry11}. Moreover, the photons that interfere at the HA stage of our setup are generated by two different photon-pair sources. The use of photons from an independent source for $\ket{\psi_{\rm av}}$ is motivated by the fact that this photon is used on the other side from the loss, which in a field-deployed application would represent a long optical channel, and thus it cannot be generated by the same source as the resource states $\ket{\psi_{\rm fe}}$ and $\ket{\psi_{\rm hg}}$. Quantum (Hong-Ou-Mandel or HOM) interference of photons from independent sources requires photons that are spectrally pure and indistinguishable. The conventional approach is to apply lossy spectral filtering. In this case, the extra loss does not not only reduce the efficiency of the HA stage, but also increases the negative effect of high-photon number noise on the error-correction scheme. Such noise appears due to the events when more than one photon pair is generated from the same source at the same time. In our case, this would have reduced HOM interference visibility at the ES and HA stages, leading to false-positive heralding events and directly affecting the final state in terms of the amount of entanglement it carries. HOM interference visibilities and heralding efficiencies provided by photon source technology employed in previous works including Ref.~\cite{kocsis13} are insufficient to demonstrate the advantage of our scheme withouth relying on postselection, which was enabled by the high-performance source and detector configuration that we describe next.
	
	\subsection*{Experimental details}
	Our heralded  single-photon sources~\cite{weston16} consisted of two $2~{\rm mm}$ nonlinear periodically poled potassium titanyl phosphate crystal (pp-KTP, poling period $46.20~{\rm \mu m}$), phase-matched for type-II collinear operation, pumped by a mode-locked Ti:sapphire laser with $\approx81~{\rm MHz}$ repetition rate, $775~{\rm nm}$ wavelength and $6~{\rm nm}$ full-width at half-maximum (FWHM) bandwidth. SPDC from the pp-KTP crystals produced degenerate photon pairs centred at $1,550~{\rm nm}$ wavelength and with $\approx15~{\rm nm}$  FWHM bandwidth. The pump and collection beamwaist sizes were set to $170~{\rm \mu m}$ and $50~{\rm \mu m}$, respectively, in order to optimize the output mode matching to an optical single mode fiber. Connecting our sources directly to high-efficiency SNSPDs~\cite{marsili13}, we achieved symmetric heralding efficiencies~\cite{klyshko80} of $\approx0.8$. In the experimental setup, the observed delivery efficiency of $\ket{\psi_{\rm av}}$, including all the optical losses and non-unit detection efficiency, ranged from $0.635\pm0.004$ to $0.757\pm0.004$, depending on the particular output path towards detectors $D_1,D_2,D_+,D_-$. The lower efficiency (compared to the value for source and detectors only) is due to the delivery efficiency which includes non-unit transmission of polarizing beam splitters (PBS) inside the HA and measurement stages and  different efficiency of the SNSPDs at different outputs. 
	
	The GVM SPDC sources generate photon pairs with high intrinsic spectral purity, providing HOM interference visibility up to $0.84$ between photons from independent sources without applying any spectral filtering. This number can be boosted up to almost one by applying mild spectral filtering only on the corresponding herald photons~\cite{weston16}, leaving interfering photons unfiltered and thus not affected by extra loss. Recent advances in design and engineering of poling of nonlinear crystals were able to provide photon-pair sources that require no filtering at all to achieve almost unit HOM interference visibility between photons from different sources~\cite{pickston21}. Our work did not rely on such poling engineering and we achieved high-visibility interference by applying bandpass filters with $\approx 8~{\rm nm}$ FWHM on the herald of $\ket{\psi_{\rm av}}$ and on the photons that propagate inside the ES stage of the setup, as shown in Fig.~\ref{fig2}{\bf a}. No spectral filtering was applied on the photons that interfered at the HA stage. In such configuration, we observe independent (photons from different sources) HOM interference visibility of up to $0.97\pm0.03$ at the HA stage and dependent (photons from the same source) HOM interference visibility of $0.990\pm0.001$ at the ES stage, see Supplementary Figure {\bf 1}.
	
	The pump power of the sources was set so to achieve pair-per-pulse generation probability of $p\approx0.00123$, equivalent to a probability of $p^2\approx1.5\times 10^{-6}$ of generating two photon pairs from the same source.
	
	Polarization mode manipulation was implemented via PBS and half- and quarter-wave plates (HWP and QWP). Polarization-dependent loss was implemented with a polarizing beam splitter (PBS), rotated around its vertical axis away from the normal incidence condition, accompanied by a HWP for phase and walk-off compensation. 
	
	\subsection*{Data availability}
	The data that support the plots within this paper and other findings of this study are available from  corresponding authors upon reasonable request.
\section{References}	
%

\section{Acknowledgements}
	This work was conducted by the ARC Centre of Excellence for Quantum Computation and Communication Technology under grant CE170100012. M.M.W. acknowledges financial support through Australian Government Research Training Program Scholarships.
	\vspace{1 EM}
	
\section{Author contributions}
	
	\noindent T.C.R. and G.J.P. conceived the idea and supervised the project. S.S. and M.M.W. developed the theoretical model with help from T.C.R., and constructed and carried out the experiment with help from L.K.S., S.K. and G.J.P. V.B.V. and S.-W.N. developed the high-efficiency SNSPDs. All authors discussed the results and contributed to the manuscript. Correspondence and requests for materials should be addressed to S.S., T.C.R., and G.J.P.
	
\section{Competing  interests}
	
	\noindent The authors declare no competing interests.

\setcounter{figure}{0}
\setcounter{equation}{0}

\def\bibsection{\subsection*{Supplementary References}} 
\renewcommand{\figurename}{{\bf Supplementary Figure}}
\renewcommand{\tablename}{{\bf Supplementary Table}}
\renewcommand{\thefigure}{{\bf \arabic{figure}}}
\renewcommand{\thetable}{{\bf \Roman{table}}}
\renewcommand{\theequation}{S\arabic{equation}} 
\makeatletter
\renewcommand{\@caption@fignum@sep}{\ ${\bm |}$\ }%
\makeatother

\onecolumngrid
\section{Supplementary Information \\for\\Quantum channel correction outperforming direct transmission}

\subsection{Supplementary Section 1: Source Data}

\begin{table}[h]
		\caption{{\bf Concurrence measurements $\bm{C_{\rm hv}}$ of entanglement distributed via the error-corrected quantum channel as a function of $\bm\eta$.} {\bf a} Data for the case of $L=0.9884\pm0.0006$, {\bf b} Data for the case of $L=0.958\pm0.001$, {\bf c} Data for the case of $L=0.903\pm 0.002 $. Reported uncertainties $\Delta C$ correspond to the experimentally observed statistical uncertainty of $\pm1$ standard deviation. Green (red) background highlights the cases when the observed concurrence was higher (lower) that the concurrence $C_{\rm fe}$ of the state distributed directly through loss, taking into account the uncertainties. $C_{\rm fe} = 0.065\pm 0.001$,$0.121\pm0.002$, and $0.184\pm0.002$ for {\bf a}, {\bf b}, and {\bf c}, respectively.  For the gain setting $\eta=0.0049$ and loss $L=0.9884$ (i.e.\ the results highlighted in bold in \textbf{a}), the concurrence is estimated theoretically to be increased to $\approx 0.52$, under the assumptions of perfect HOM interference, a fully-lossless setup (other than loss $L$ on the channel), unit-efficiency threshold detection, but taking into account high-order photon number noise. The expected direct transmission concurrence for the same experimental conditions is estimated to be $\approx0.1$.}
			
	\begin{tabular}{|C{1.2cm}|C{1cm}|C{1cm}|}

		\hline
		\multicolumn{3}{|c|}{{\bf a}\quad $L=0.9884\pm0.0006$}\\
		\hline
		\hline
		$\eta$&  $ C_{\text{hv}}$ &    $\Delta C_{\text{hv}}$ \\
		\hline
		\hline
\rowcolor{goodcolor}	$	0.0012$&   $ 0.22   $&$ 0.03$ \\
\rowcolor{goodcolor}	$	{\bf 0.0049}$& $ {\bf 0.27}   $&$ {\bf 0.02}$ \\
\rowcolor{goodcolor}	$	0.0076$&   $ 0.23   $&$ 0.02$ \\
\rowcolor{goodcolor}	$	0.0302$&   $ 0.14   $&$ 0.02$ \\
\rowcolor{goodcolor}	$	0.0670$&   $ 0.08   $&$ 0.01$ \\
		\hline
\rowcolor{badcolor}	$	0.1170$&   $ 0.06$&   $ 0.01$ \\
\rowcolor{badcolor}	$	0.1786$&  $ 0.033$&   $ 0.006$ \\
\rowcolor{badcolor}	$	0.2500$&   $ 0.022$&   $ 0.007$ \\
		\hline
	\end{tabular}\quad
	\begin{tabular}{|C{1.2cm}|C{1cm}|C{1cm}|}
		\hline
		\multicolumn{3}{|c|}{{\bf b}\quad $L=0.958\pm0.001$}\\
		\hline
		\hline
		$\eta   $& $ C_{\text{hv}} $& $ \Delta C_{\text{hv}}$ \\
		\hline
		\hline
\rowcolor{badcolor}	$ 0.0012 $&  $ 0.13  $& $ 0.03 $\\
		\hline
\rowcolor{goodcolor}$	0.0049 $&  $ 0.21  $& $ 0.03$ \\
\rowcolor{goodcolor}$	0.0076 $&  $ 0.23  $& $ 0.04$ \\
\rowcolor{goodcolor}$	0.0302 $&  $ 0.23  $& $ 0.01$ \\
\rowcolor{goodcolor}$	0.0670 $&  $ 0.16  $& $ 0.02$ \\
		\hline
\rowcolor{badcolor}$	0.1170 $&  $ 0.13  $& $ 0.01$ \\
\rowcolor{badcolor}$	0.1786 $&  $ 0.09  $& $ 0.01$ \\
\rowcolor{badcolor}$	0.2500 $&  $ 0.08  $& $ 0.01$ \\
		\hline
	\end{tabular}\quad
	\begin{tabular}{|C{1.2cm}|C{1cm}|C{1cm}|}
		\hline
		\multicolumn{3}{|c|}{{\bf c}\quad $L=0.903\pm 0.002$ }\\
		\hline
		\hline
		$\eta  $&  $ C_{\text{hv}} $&  $ \Delta C_{\text{hv}}$ \\
		\hline
		\hline
\rowcolor{badcolor}$	0.0012  $& $ 0.10  $& $ 0.03$ \\
\rowcolor{badcolor}$	0.0049  $& $ 0.17  $& $ 0.03$ \\
\rowcolor{badcolor}$	0.0076  $& $ 0.22  $& $ 0.04$ \\
		\hline
\rowcolor{goodcolor}$	0.0302  $& $ 0.25  $& $ 0.01$ \\
\rowcolor{goodcolor}$	0.0670  $& $ 0.22  $& $ 0.02$ \\
		\hline
\rowcolor{badcolor}$	0.1170  $& $ 0.17  $& $ 0.01$ \\
\rowcolor{badcolor}$	0.1786  $& $ 0.13  $& $ 0.01$ \\
\rowcolor{badcolor}$	0.2500  $& $ 0.12  $& $ 0.01$ \\
		\hline
	\end{tabular}

\end{table}

\begin{table}[h]
		\caption{{\bf Absolute values of the density matrix elements (Fig.~{\bf 4a}, {\bf c}, {\bf e}) of the state $\bm{\rho_{\rm fe}}$ transmitted directly through loss.} {\bf a} Data for the case of $L=0.9884\pm0.0006$, {\bf b} Data for the case of $L=0.958\pm0.001$, {\bf c} Data for the case of $L=0.903\pm 0.002 $.
	}
\parbox{0.32\linewidth}{
	\centering {\bf a}\qquad $L=0.9884\pm0.0006$
		\begin{tabular}{|C{1cm}|C{1cm}C{1cm}C{1cm}C{1cm}|}
				&	$\bra{0,0}  $& $ \bra{1,0}  $& $ \bra{0,1}  $&  $\bra{1,1}$\\
		\hline
		$\ket{0,0}$& $	0.67376   $&  $ 0			$&  $ 0			   $& $ 0$ \\
		$\ket{1,0}$& $	0		  $&  $ 0.32113		$&  $ 0.04016	   $& $ 0$ \\
		$\ket{0,1}$& $	0		  $&  $ 0.04016		$&  $ 0.00502	   $& $ 0$ \\
		$\ket{1,1}$& $	0		  $&  $ 0			$&  $ 0	  		   $& $ 0.00008$ \\
		\end{tabular}
}	
\parbox{0.32\linewidth}{
	\centering {\bf b}\qquad $L=0.958\pm0.001$
		\begin{tabular}{|C{1cm}|C{1cm}C{1cm}C{1cm}C{1cm}|}
				&	$ \bra{0,0}  $& $ \bra{1,0} $&  $ \bra{0,1} $&  $ \bra{1,1}$\\
		\hline
		$\ket{0,0}$& $	0.68105	 $&   $ 0			$&  $ 0			   $& $ 0$ \\
		$\ket{1,0}$& $	0		 $&   $ 0.30296	$&	  $ 0.06935	   $& $ 0$ \\
		$\ket{0,1}$& $	0		 $&   $ 0.06935	$&	  $ 0.01587	   $& $ 0$ \\
		$\ket{1,1}$& $	0		 $&   $ 0			$&  $ 0			   $& $ 0.00011$ \\
		\end{tabular}
}
\parbox{0.32\linewidth}{
	\centering {\bf c}\qquad $L=0.903\pm 0.002 $	
		\begin{tabular}{|C{1cm}|C{1cm}C{1cm}C{1cm}C{1cm}|}
					& $  \bra{0,0}$&  $\bra{1,0}$&  $\bra{0,1}$&$  \bra{1,1}$\\
		\hline
		$\ket{0,0}$& $ 0.66403	   $& $ 0			 $& $ 0			   $& $ 0$ \\
		$\ket{1,0}$& $ 0			$&    $ 0.30447	$&	  $ 0.09784	   $& $ 0$ \\
		$\ket{0,1}$& $ 0			$&    $ 0.09784	$&	  $ 0.03144	   $& $ 0$ \\
		$\ket{1,1}$& $ 0			$&    $ 0		$&	  $ 0			$&    $ 0.00005$ \\
		\end{tabular}
}
\end{table}

\begin{table}[h]
	\caption{{\bf Absolute values of the density matrix elements (Fig.~{\bf 4b}, {\bf d}, {\bf f}) of the state $\bm{ \rho_{\rm hv}}$ transmitted through the error corrected channel.} {\bf a} Data for the case of $L=0.9884\pm0.0006$, {\bf b} Data for the case of $L=0.958\pm0.001$, {\bf c} Data for the case of $L=0.903\pm 0.002 $.
	}
\parbox{0.32\linewidth}{
	\centering {\bf a}\qquad $L=0.9884\pm0.0006$
		\begin{tabular}{|C{1cm}|C{1cm}C{1cm}C{1cm}C{1cm}|}
					& $ \bra{0,0} $&$ \bra{1,0} $&$ \bra{0,1} $&$ \bra{1,1}$\\
		\hline
		$\ket{0,0} $& $ 0.67284	 $&   $ 0			$&  $ 0 		  $&  $ 0$ \\
		$\ket{1,0} $& $ 0		$&	    $ 0.11722	$&	  $ 0.13349 	$&    $ 0$ \\
		$\ket{0,1} $& $ 0		$&	    $ 0.13349	$&	  $ 0.20994 	$&    $ 0$ \\
		$\ket{1,1} $& $ 0		$&	    $ 0			$&  $ 0			 $&   $ 0$ \\
		\end{tabular}
}	
\parbox{0.32\linewidth}{
	\centering {\bf b}\qquad $L=0.958\pm0.001$
		\begin{tabular}{|C{1cm}|C{1cm}C{1cm}C{1cm}C{1cm}|}
					&$  \bra{0,0} $&$ \bra{1,0} $&$ \bra{0,1} $&$ \bra{1,1}$\\
		\hline
		$\ket{0,0}$&  $ 0.52014	$&    $ 0 			$&  $ 0			  $&  $ 0$ \\
		$\ket{1,0} $& $ 0		$&	    $ 0.06537	$&	  $ 0.13310	  $&  $ 0$ \\
		$\ket{0,1} $& $ 0		$&	    $ 0.13310	$&	  $ 0.41401	  $&  $ 0$ \\
		$\ket{1,1} $& $ 0		$&	    $ 0			$&  $ 0 		  $&  $ 0.00049$ \\
		\end{tabular}
}	
\parbox{0.32\linewidth}{
	\centering {\bf c}\qquad $L=0.903\pm 0.002 $	
		\begin{tabular}{|C{1cm}|C{1cm}C{1cm}C{1cm}C{1cm}|}
					&$  \bra{0,0} $&$ \bra{1,0} $&$ \bra{0,1} $&$ \bra{1,1}$\\
		\hline
		$\ket{0,0}$&  $ 0.58178	  $&  $ 0			$&  $ 0			  $&  $ 0$ \\
		$\ket{1,0}$&  $ 0		$&	    $ 0.08891 	$&  $ 0.14157 	  $&  $ 0$ \\
		$\ket{0,1}$&  $ 0		$&	    $ 0.14157	$&	  $ 0.32895	  $&  $ 0$ \\
		$\ket{1,1}$&  $ 0		$&	    $ 0			$&  $ 0			  $&  $ 0.00037$ \\
		\end{tabular}
}	
\end{table}

\subsection{Supplementary Section 2: Heralded amplification background theory}

The effect of loss on a single rail qubit is equivalent to an amplitude damping channel. In particular the pure initial state $\alpha |0 \rangle + \beta |1 \rangle$ is taken to:
\begin{equation}
|\beta|^2 (1- T) |0\rangle\langle 0| + (\alpha |0 \rangle +\sqrt{T} \beta |1 \rangle)(\alpha^* \langle 0| + \sqrt{T}\beta^*\langle 1|)
\end{equation}
The successful action of the NLA with gain $g$ on this state gives:
\begin{equation}
{{|\beta|^2 (1- T) |0\rangle\langle 0| + (\alpha |0 \rangle + g \sqrt{T} \beta |1 \rangle)(\alpha^* \langle 0| + g \sqrt{T} \beta^*\langle 1|)}\over{1+T |\beta|^2(g^2-1)}}
\end{equation}
The fidelity between the initial and final states is:
\begin{equation}
F={{|\beta|^2 (1- T) (1-|\beta|^2)+ (1-|\beta|^2+g \sqrt{T} |\beta|^2)^2}\over{1+T |\beta|^2(g^2-1)}}
\end{equation}
The fidelity is improved by successful action of the NLA by choosing an optimal value of the gain. However, the optimal gain is state dependent (and $T$ dependent). To remove the state dependence we can consider the average fidelity over the Block sphere, given by:
\begin{equation}
F_{av}=\int_{0}^{1} F d|\beta|^2
\end{equation}
which still shows an advantage from use of the NLA with an optimized gain. The disadvantages of this approach are that the improvement is limited and the state is destroyed when the NLA doesn't succeed.

An alternative approach is to send one arm of the entangled state $\sqrt{\epsilon} |01 \rangle + \sqrt{1-\epsilon} |10 \rangle$ through the channel. After the loss and successful application of the NLA with the gain $g \sqrt{\epsilon T} = \sqrt{1-\epsilon}$ the state is:
\begin{equation}
{{\epsilon (1- T) |00\rangle\langle 00| + (1-\epsilon)(|01 \rangle + |10 \rangle)( \langle 01| + \langle 10|)}\over{2(1-\epsilon)+(1-T) \epsilon}}
\end{equation}
In the limit that $\epsilon$ is small one approximately retrieves the maximally entangled state ${{1}\over{\sqrt{2}}}((|01 \rangle + |10 \rangle)$. In principle this state can be then be used to teleport arbitrary states forming an effective identity channel in spite of arbitrary loss on the physical channel. In practice the feedforward gain becomes very large in this limit and hence the probability of preparing the state becomes very low making this regime difficult to achieve. In the experiment we use equal superposition states ($\epsilon = 1/2$) and find that for a range of gains we are able to demonstrate error corrected channels which outperform the direct channel by a large margin.

\subsection{Supplementary Section 3: Protocol rates data and discussion}

A fair comparison between the direct transmission and error-corrected channel performance requires precise accounting of the operation rates of the two channels. Due to the probabilistic nature of SPDC, the probability of generating two photon pairs for the full protocol is lower than the probability of generating a single pair for the direct transmission. This, together with the fact that HA success rate decreases with the increase of gain setting, leads to the lower absolute counts per second for the error corrected channel compared to the direct transmission.

In our analysis, however, we use a more relevant definition of the operation rate as the rate at which input states are prepared and sent through the channel, given successful preparation of the channel. For the error-corrected channel, the preparation of the channel is heralded by the joint heralding signal from HA stage and the herald of the ancilla photon. The success of the input state preparation and transmission through the channel is heralded by the joint signal from the ES, HA and the ancilla herald detection. The corresponding experimental heralding rates are shown in Supplementary Tables~\ref{table:channel} and~\ref{table:stetaandchannel} for all three values of added loss and different gain settings of the HA. The operation rate of the direct transmission channel is defined as the ratio between the the rate of state heralding and the pulse rate of the laser. Supplementary Figure~\ref{fig:ratios} shows the comparison between operation rates for the two types of channels.

For low amplifier gain and low amplitude of the state input to HA, the probability of success of the amplifier is approximately independent of the input state. In this regime the firing of the entanglement swapping detectors and the noiseless amplification detectors are independent events and we see the true rate of entanglement swapping normalized to the random rate of noiseless amplification successes. In this regime we see equal rates for direct transmission and error-corrected transmission. The use of photon number resolving detectors at the entanglement swapping stage would lead to the $50\%$ failure rate of the teleported channel, which would need to be taken into account for the correct comparison of the concurrences. Instead, the use of threshold detectors, although decreasing the final concurrence of the state sent through the error-corrected channel, provides a level playing field for the comparison of the two scenarios.

When the gain of the amplification is high, i.e. $\eta\approx0$, the probability of the HA resrource state $\ket{\psi_{\rm av}}=\sqrt{\eta}\ket{1_a 0_v}+\sqrt{1-\eta}\ket{0_a 1_v}$ to be transmitted inside HA stage becomes comparable with the probability of receiving the entangled resource state $\ket{\psi_{\rm fe}}$. The success of amplification thus becomes affected by the amplitude of  $\ket{\psi_{\rm fe}}$ sent through the channel. As the preparation of the input state $\ket{\psi_{\rm hg}}$ and the preparation of the entangled state $\ket{\psi_{\rm fe}}$ are a correlated event for our particular setup, the success of the swapping and the amplification become correlated too. This leads to the rate in the case of the error corrected channel being higher than direct transmission. However, given this is a peculiarity of our specific set-up, we consider the true rate to be that determined in the low gain limit. Given then equal rates, the direct comparison of the concurrences in the two cases is justified. 

\begin{table}[h!]
	\caption{\label{table:channel}{\bf Error-corrected channel preparation rate.} The rates of  heralding signal from HA stage, measured in Hz, as a function of $\eta$, for three different values of added loss $L$.}
	\begin{tabular}{|C{2.8cm}||C{1.75cm}|C{1.75cm}|C{1.75cm}|C{1.75cm}|C{1.75cm}|C{1.75cm}|C{1.75cm}|C{1.75cm}|}
	\hline
		$\eta$  &  $ 0.0012 $&  $ 0.0049 $&  $ 0.0076 $& $ 0.0302  $& $ 0.0670 $&  $ 0.1170  $& $ 0.1786  $& $ 0.2500$ \\
		\hline
		\hline

$L = 0.9884 \pm  0.0006 $&$ 76.75\pm0.07 $ & $ 165.6\pm0.1 $ & $ 242.3\pm0.1 $ & $ 988.9\pm0.6 $ & $ 1996.1\pm0.3 $ & $ 3587.6\pm0.5 $ & $ 4411.0\pm0.5 $ & $ 7303.1\pm0.8 $ \\
$L = 0.958 \pm  0.001 $&$ 50.22\pm0.05 $ & $ 113.82\pm0.09 $ & $ 160.4\pm0.1 $ & $ 580.4\pm0.2 $ & $ 1329.3\pm0.3 $ & $ 2540.1\pm0.6 $ & $ 4026.8\pm0.7 $ & $ 5654.9\pm0.8 $ \\
$L = 0.903 \pm  0.002  $&$ 49.40\pm0.06 $ & $ 85.42\pm0.08 $ & $ 150.9\pm0.1$ & $ 478.3\pm0.2 $ & $ 1243.2\pm0.4 $ & $ 2542.1\pm0.5 $ & $ 3969.7\pm0.6 $ & $ 5604.0\pm0.8 $ \\
		\hline
	\end{tabular}
\end{table}

\begin{table}[h]
	\caption{\label{table:stetaandchannel} {\bf Error-corrected channel state preparation and sending  rates.} The rates of joint heralding signal from ES and HA stages, measured in Hz, as a function of $\eta$, for three different values of added loss $L$.}
	\begin{tabular}{|C{2.8cm}||C{1.75cm}|C{1.75cm}|C{1.75cm}|C{1.75cm}|C{1.75cm}|C{1.75cm}|C{1.75cm}|C{1.75cm}|}
	\hline
	$\eta$  &  $ 0.0012 $&  $ 0.0049 $&  $ 0.0076 $& $ 0.0302  $& $ 0.0670 $&  $ 0.1170  $& $ 0.1786  $& $ 0.2500$ \\
		\hline
	\hline
	$L = 0.9884 \pm  0.0006 $& $ 0.062 \pm 0.002  $& $ 0.099 \pm 0.002  $& $ 0.127 \pm 0.003  $& $ 0.43 \pm 0.01  $& $ 0.824 \pm 0.007  $& $ 1.47 \pm 0.01  $& $ 1.77 \pm 0.01  $& $ 2.99 \pm 0.02 $\\
	$L = 0.958 \pm  0.001 $& $ 0.101 \pm 0.002  $& $ 0.117 \pm 0.003 $& $ 0.133 \pm 0.003  $& $ 0.281 \pm 0.005  $& $ 0.524 \pm 0.007  $& $ 0.94 \pm 0.01  $& $ 1.47 \pm 0.01  $& $ 2.00 \pm 0.01 $\\
	$L = 0.903 \pm  0.002  $&$ 0.183 \pm 0.004  $& $ 0.158 \pm 0.004  $& $ 0.207 \pm 0.005  $& $ 0.282 \pm 0.005  $& $ 0.499 \pm 0.007  $& $ 1.03 \pm 0.01  $& $ 1.50 \pm 0.01  $& $ 2.04 \pm 0.01$ \\
	\hline
	\end{tabular}
	
\end{table}
	\begin{figure}[h]
		\includegraphics{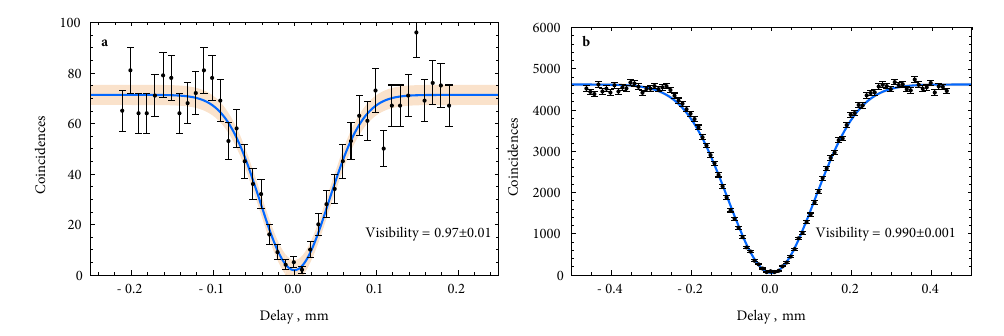}
		\caption{\label{fig:HOM} {\bf Hong-Ou-Mandel interference.} {\bf a} Four-photon coincidence rate of a HOM interference at the HA stage as a function of an optical delay introduced between the interfering photons. Mild, $\approx8~\rm{nm}$ FWHM, spectral filtering applied on the herald photons only. {\bf b} Two-photon coincidence rate of a HOM interference at the ES stage. Same $\approx8~\rm{nm}$ FWHM spectral filtering is applied to interfering photons, resulting only in the widening of the HOM dip, with minimal effect on interference visibility. Error bars correspond to the experimentally observed statistical uncertainty of $\pm1$ standard deviation. Shaded areas correspond to the $95\%$ confidence region, derived from uncertainty in the fit parameters.}
	\end{figure}
	
	\begin{figure}[h]
	\includegraphics{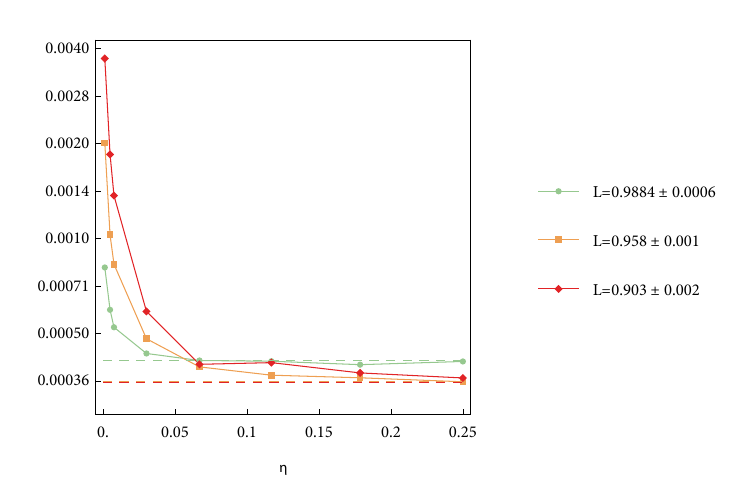}
		\caption{\label{fig:ratios} {\bf Logarithmic plot of the operating rates of the error-corrected and direct transmission quantum channels for three different values of added loss.} Dashed lines represent the experimentally measured operating rate of the direct transmission through loss, and dots represent the operating rates of the error-corrected channel as a function of the HA gain setting. Error bars are smaller than the dots size.}
\end{figure}
 
\end{document}